\documentclass[aps,prb,twocolumn,superscriptaddress,showpacs,10pt]{revtex4-1}
\usepackage{graphicx}
\usepackage{calc}
\usepackage{bm}
\usepackage{color}

\begin{document}

\title{Evidence on the macroscopic length scale spin coherence for the edge currents in a narrow HgTe quantum well}

\author{A.~Kononov}
\affiliation{Institute of Solid State Physics RAS, 142432 Chernogolovka, Russia}
\author{S.V.~Egorov}
\affiliation{Institute of Solid State Physics RAS, 142432 Chernogolovka, Russia}
\author{Z.D. Kvon}
\affiliation{Institute of Semiconductor Physics, Novosibirsk 630090, Russia}
\affiliation{Novosibirsk State University, Novosibirsk 630090, Russia}
\author{N.N. Mikhailov}
\affiliation{Institute of Semiconductor Physics, Novosibirsk 630090, Russia}
\author{S.A. Dvoretsky}
\affiliation{Institute of Semiconductor Physics, Novosibirsk 630090, Russia}
\author{E.V.~Deviatov}
\affiliation{Institute of Solid State Physics RAS, 142432 Chernogolovka, Russia}

\date{\today}

\begin{abstract}
We  experimentally  investigate spin-polarized electron transport between two ferromagnetic contacts, placed at the edge of a two-dimensional electron system with band inversion. The system is realized in a narrow (8~nm) HgTe quantum well, the ferromagnetic side contacts are formed from a pre-magnetized permalloy film. In zero magnetic field,  we find a significant edge current contribution to the transport between two ferromagnetic contacts. We experimentally demonstrate that this transport is sensitive to the mutual orientation of the magnetization directions of two 200~$\mu$m-spaced ferromagnetic leads. This is a direct experimental evidence on the spin-coherent edge transport over the macroscopic  distances. Thus, the spin is extremely robust at the edge of a two-dimensional electron system with band inversion, confirming the helical spin-resolved nature of edge currents.
\end{abstract}

\pacs{73.40.Qv  71.30.+h}

\maketitle

%\section{Introduction}

Recently, there is a strong interest in two-dimensional semiconductor systems with  band inversion, like narrow HgTe quantum wells. This interest is mostly connected with the quantum spin-Hall effect (QSHE) regime~\cite{konig,kvon}  in zero magnetic field. Similarly to the conventional quantum Hall (QH) effect in high magnetic fields~\cite{buttiker}, QSHE  is characterized~\cite{molenkamp_nonlocal,kvon_nonlocal} by edge state transport. In contrast to the  chiral~\cite{buttiker} transport in the QH regime, these QSHE edge states are helical, i.e. two spin-resolved edge states are counter-propagating at a particular sample edge~\cite{pankratov,zhang1,kane,zhang2}. Experimental investigation of helical edge states is based on the charge transport along the sample edge, which has been detected in local and  non-local resistance measurements~\cite{konig,kvon,molenkamp_nonlocal,kvon_nonlocal} and by a direct visualization technique~\cite{imaging}.  

The helical edge transport has to be essentially spin-dependent. Two spin-resolved counter-propagating edge states are  supposed to be responsible for the topological protection, which is a key feature of a topological isolator regime.~\cite{pankratov,zhang1,kane,zhang2}  Some signature of the spin transport in QSHE edge states was demonstrated by means of metallic spin Hall transport in nanoscale structures~\cite{molenkamp_spin}. On the other hand, spin effects are supposed~\cite{halperin06,qi} to be mostly prominent for the semiconductor-ferromagnet hybrid structures,  where the ferromagnetic leads allow the possibility of spin-polarized current injection and/or detection at the sample edge~\cite{feinas}.

Here, we experimentally  investigate spin-polarized electron transport between two ferromagnetic contacts, placed at the edge of a two-dimensional electron system with band inversion. The system is realized in a narrow (8~nm) HgTe quantum well, the ferromagnetic side contacts are formed from a pre-magnetized permalloy film. In zero magnetic field,  we find a significant edge current contribution to the transport between two ferromagnetic contacts. We experimentally demonstrate that this transport is sensitive to the mutual orientation of the magnetization directions of two 200~$\mu$m-spaced ferromagnetic leads. This is a direct experimental evidence on the spin-coherent edge transport over the macroscopic  distances. Thus, the spin is extremely robust at the edge of a two-dimensional electron system with band inversion, confirming the helical spin-resolved nature of edge currents.

%\section{Samples and technique}

\begin{figure}
\includegraphics[width=\columnwidth]{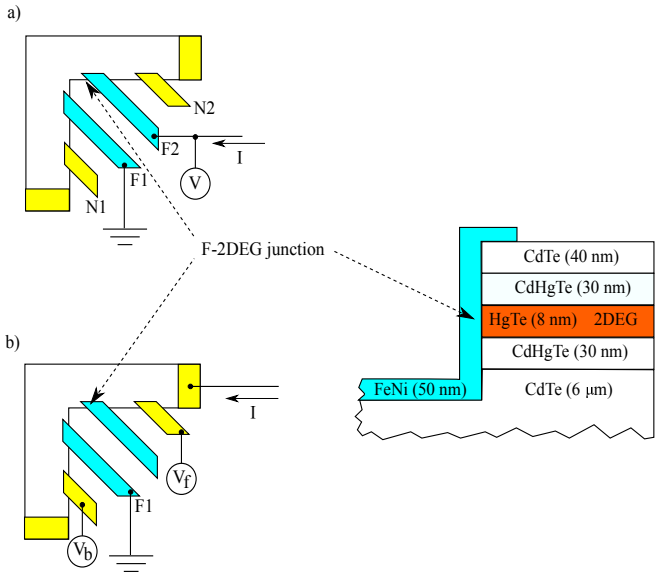}
\caption{(Color online) Sketch of the sample (not in scale) with electrical connections in different experimental configurations. The 100~$\mu$m wide corner-shape mesa is formed by dry etching (200 nm deep) in Ar plasma. Two ferromagnetic $Fe_{20}Ni_{80}$ permalloy stripes (blue, F1 and F2) are placed at the mesa step, with low (2-3~$\mu$m) overlap. In every overlap region, a side F-2DEG junction is formed  between the ferromagnetic  lead and the 2DEG edge. The width of each junction is equal to 20~$\mu$m. The junctions F1 and F2 are separated by the 300~$\mu$m distance along the sample edge. Several Au stripes (yellow) are placed at the mesa step, to form several normal (N) reference junctions (only N1 and N2 are shown).  We use a standard two-point F-2DEG-F experimental configuration (a), realized by grounding the ferromagnetic lead F2  and using F1 to apply a current and to measure a voltage drop simultaneously. We also study electron transport across one particular F-2DEG junction (b): the corresponding ferromagnetic electrode is grounded (F1); a current is applied between it and one of the  normal contacts; two other contacts trace the 2DEG potential to both sides of the grounded junction, $V_f$  and $V_b$, respectively. }
\label{sample}
\end{figure}

Our $Cd_{0.65}Hg_{0.35}Te/HgTe/Cd_{0.65}Hg_{0.35}Te$ quantum wells with [013] surface orientations and width $d$ of $8$-$8.3$~nm are grown by molecular beam epitaxy, see Fig.~\ref{sample}. A detailed description of the sample structure is given elsewhere~\cite{growth1,growth2}. Because of $d$ above the critical value 6.3~nm, the quantum wells are characterized by band inversion~\cite{kvon,kvon_nonlocal}. They contain a two-dimensional electron gas (2DEG) with the electron  density  of  $1.5 \cdot 10^{11}  $cm$^{-2}$, as obtained from standard magnetoresistance measurements. The 2DEG mobility at 4K equals to $2\cdot 10^{5}  $cm$^{2}$/Vs.

A sample sketch is presented in Fig.~\ref{sample} (a) and (b). The 100~$\mu$m wide corner-shape mesa is formed by dry etching (200 nm deep) in Ar plasma. We fabricate F-2DEG junctions by using rf sputtering to deposit 30~nm thick ferromagnetic permalloy $Fe_{20}Ni_{80}$   stripes over the mesa step, with low (2-3~$\mu$m) overlap.  The stripes are formed by lift-off technique, and the surface is mildly cleaned by Ar plasma before sputtering.  To avoid any 2DEG  degradation, the sample is not  heated during the sputtering process. The reference junctions N1 and N2 and the source-drain contacts  are obtained by thermal evaporation of 100~nm thick Au (yellow in Fig.~\ref{sample}). Without annealing procedure, only a side contact is possible between the metallic electrode and the 2DEG edge at the mesa step, because of the insulating CdTe layer on the top of the structure.

We use a standard two-point F-2DEG-F experimental configuration, realized  by grounding one ferromagnetic lead and using another one to apply a current and to measure a voltage drop simultaneously, see Fig.~\ref{sample} (a). We also study electron transport across one particular F-2DEG junction, see Fig.~\ref{sample} (b): the corresponding ferromagnetic electrode is grounded; a current is applied between it and one of the  normal contacts; two other contacts trace the 2DEG potential to both sides  of the grounded junction, $V_f$  and $V_b$, respectively. 

\begin{figure}
\includegraphics[width=\columnwidth]{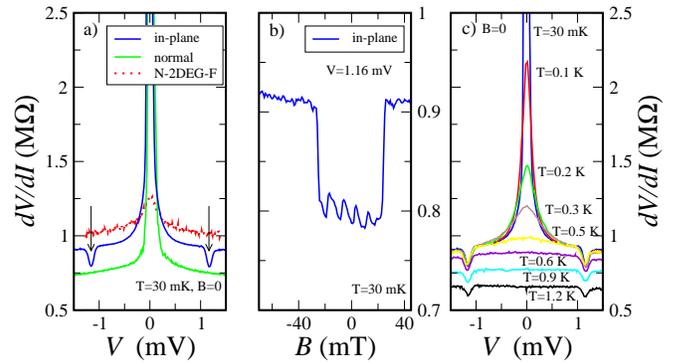}
\caption{(Color online) (a) Two-point differential resistance $dV/dI(V)$ between two ferromagnetic leads (F1-2DEG-F2 junction) for two (normal and in-plane) permalloy magnetizations. We observe strong and narrow deeps in differential resistance, placed at $\pm 1.16$~mV bias, for the permalloy film magnetization oriented within the 2DEG plane (blue line). These deeps are not seen for the normal magnetization orientation (green line), or if any of two ferromagnetic leads is changed to a normal one (N-2DEG-F configuration, dots). (b) The external magnetic field sharply suppresses the resistance deeps at $B=\pm 25$~mT. Between $B=\pm 25$~mT, $dV/dI(B)$ curve demonstrates  well-reproducible oscillations with a period $\Delta B \approx 10$~mT. (c) Temperature dependence of the two-point $dV/dI(V)$ F1-2DEG-F2 curve for the in-plane permalloy magnetization in zero magnetic field.  The conductance peaks at $\pm 1.16$~mV bias are only weakly sensitive to the temperature below 1~K. 
}
\label{IVgate}
\end{figure}

\begin{figure}
\includegraphics[width=\columnwidth]{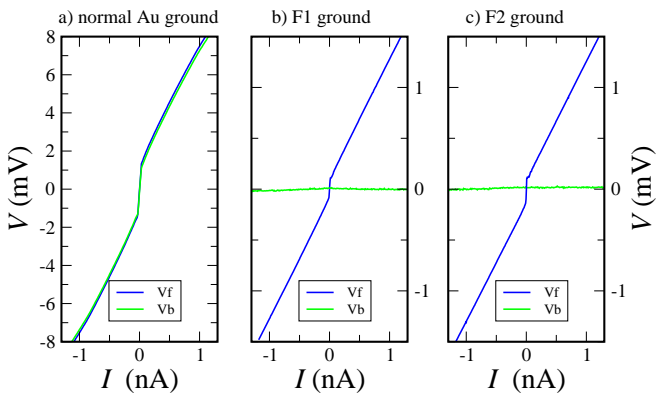}
\caption{(Color online) Examples of  $I-V$ characteristics for transport across a single N-2DEG (a) or F-2DEG (b,c) junction. The curves are obtained in the experimental configuration depicted in Fig.~\protect\ref{sample} (b). For  a reference  Au junction N1 (a), $V_f$ and $V_b$ coincide well and reflect the resistance of the junction N1 in a standard three-point configuration. For the ferromagnetic junctions F1 (b) or F2 (c), the measured $I-V$ curve is crucially dependent on the mutual positions of current and voltage probes.  We obtain significant signal $V_f$, i.e. for the voltage probe placed between the current and ground ones, but $V_b$ is always zero. The measurements are performed at a temperature of 30~mK in zero magnetic field.
}
\label{IVnl}
\end{figure}

To obtain $I-V$ characteristics, depicted in Fig.~\ref{IVnl}, we sweep the dc current from -1~nA to +1~nA and measure the dc voltage in a mV range by a dc electrometer. To obtain $dV/dI(V)$ characteristics in Fig.~\ref{IVgate}, this dc current is additionally modulated by a low ac component (0.01~nA, 2~Hz). We measure the ac ($\sim dV/dI$) component of the 2DEG potential by using a lock-in with a 100~M$\Omega$ input preamplifier. We have checked, that the lock-in signal is independent of the modulation frequency in the range 1~Hz -- 6~Hz, which is defined by applied ac filters. 

The measurements are performed at a temperature of 30~mK. To realize a spin-polarized transport~\cite{feinas}, the permalloy stripes are initially pre-magnetized in the 2DEG plane. The sample is placed within a superconducting solenoid, so the initial in-plane magnetization can be changed to a normal one by introducing relatively high (above 1~T) external magnetic field. The field is switched to zero afterward, so most of the measurements are performed in zero magnetic field.  Qualitatively similar results are obtained from different samples in several cooling cycles.

%\section{Experimental results}

%\subsection{Double F-2DEG-F junction}

Fig.~\ref{IVgate} (a) demonstrates two-point (see Fig.~\ref{sample} (a)) F-2DEG-F $dV/dI(V)$ dependencies for both (normal and in-plane) permalloy magnetizations. In this case, we investigate in-series connected resistances of two F-2DEG junctions and a 2DEG region between them.   The experimental $dV/dI(V)$ dependencies in Fig.~\ref{IVgate} (a) are checked to be invariant if we exchange F1 and F2 contacts in this two-point configuration. Both curves are characterized by narrow zero-bias resistive regions and linear branches at higher biases. We do not see any effect of the magnetization direction on the zero-bias resistive region in Fig.~\ref{IVgate} (a), however, it is sharply increased in an external magnetic field above 0.2~T. Thus, we should connect this region with a shallow potential barrier at the F-2DEG interface, e.g. due to the proximity magnetization~\cite{lunde},  which  is fully suppressed by a temperature increase above 0.3~K, see Fig.~\ref{IVgate} (c).

Surprisingly, we observe sharp and narrow deeps in differential resistance $dV/dI$, placed at $\pm 1.16$~mV bias, for the permalloy film magnetization oriented within the 2DEG plane, see Fig.~\ref{IVgate} (a). These deeps disappear completely for the normal to a 2DEG permalloy magnetization orientation,  see Fig.~\ref{IVgate} (a), or if any of two ferromagnetic leads is changed to a normal one (i.e. in N-2DEG-F configuration). The external magnetic field sharply suppresses the deeps above $B=\pm 25$~mT, see Fig.~\ref{IVgate} (b).  These $dV/dI$ resistance deeps at $\pm 1.16$~mV bias are weakly sensitive to the temperature below 1~K, see Fig.~\ref{IVgate} (c).

%\section{Discussion} \label{disc}

The fact that $dV/dI$ resistance deeps are controlled by the film magnetization direction, see Fig.~\ref{IVgate} (a), and low external magnetic field, see Fig.~\ref{IVgate} (b), is a direct experimental evidence on the the spin-coherent edge transport over the macroscopic (about 200~$\mu$m)  distances in a two-dimensional electron system with band inversion. Let us argument this statement.

\begin{figure}
\includegraphics[width=0.9\columnwidth]{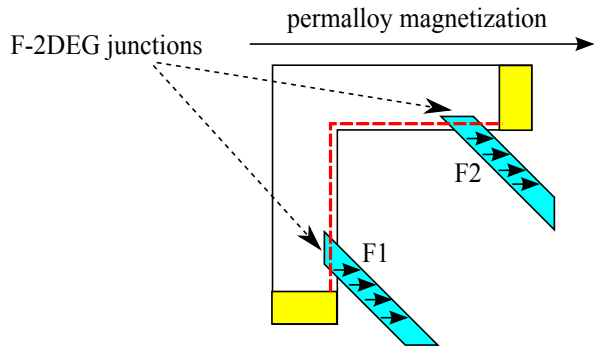}
\caption{(Color online) Schematic demonstration of the in-plane permalloy magnetization direction (arrows) and the conductive one-dimensional edge channel (red dashed line) for the junctions F1 and F2. Because of the corner-shape mesa geometry, the junctions F1 and F2 are characterized by different orientations of the permalloy magnetization and the 2DEG edge. In the figure, the magnetization is normal to the edge for the junction F1, but they are parallel for the junction F2. 
}
\label{discussion}
\end{figure}

We  observe conductance peaks only for transport between two ferromagnetic leads for the specific (in-plane) permalloy magnetization. In this case, the junctions F1 and F2 are characterized by different (normal and parallel, respectively) orientations of the permalloy magnetization to the 2DEG edge, see Fig.~\ref{discussion}, because of the corner-shape mesa geometry. In contrast, a normal to the 2DEG film magnetization is equally oriented for both the junctions F1 and F2, but the $dV/dI$ resistance deeps are not seen in this case. Also, they disappear completely  if any of two ferromagnetic junctions is changed to a normal one in Fig.~\ref{IVgate} (a) even for the in-plane magnetization. Thus, we experimentally demonstrate that transport in the present F-2DEG-F structure is sensitive to the {\em mutual} orientation of the magnetization directions at two macroscopically-spaced F-2DEG interfaces. 

This sensitivity, however, requires spin-coherent transport between the junctions F1 and F2. In this case, every F-2DEG interface is characterized by transport of spin-polarized electrons from the permalloy film to the 2DEG edge. A spin-polarized electron, emitted by the junction F1, should travel along  the sample edge to be absorbed by the junction F2. Different magnetization directions of two junctions lead to an additional potential barrier for a spin-polarized electron,  which does not occur for the equal magnetizations or for non-magnetic junction. This barrier is obviously affected by the applied bias, which should appear in $dV/dI(V)$ characteristics. Understanding of the resonance-like behavior in Fig.~\ref{IVgate} (a), however, requires a detailed analysis of the edge excitation spectrum in a narrow (8~nm) HgTe quantum well.

The spin-coherent transport can be naturally provided by the helical one-dimensional  channel~\cite{zhang1,kane,zhang2}, which consists from two spin-resolved counter-propagating edge modes. External magnetic field primary affects the helical edge modes, which results in destroying the spin coherence above $B=\pm 25$~mT  even if the field is oriented in the permalloy film magnetization direction, see Fig.~\ref{IVgate} (b). The $dV/dI$ resistance deeps weakly depend on temperature below 1~K, which is also consistent with the reported behavior of the edge state transport in QSHE regime.~\cite{kvon_temp} Another sign of the transport coherency is the well-reproducible $dV/dI$ oscillations with a constant period $\Delta B \approx 10$~mT in weak magnetic fields, which are also destroyed by the magnetic field above $B=\pm 25$~mT.  

It is important, that we observe spin coherent transport over the macroscopic distances (about 200~$\mu$m). In previous experiments on the edge transport between two non-magnetic contacts, it has been shown~\cite{konig,kvon} to be diffusive for distances above 10~$\mu$m. Thus, the spin information is much more robust at the edge of a two-dimensional electron system with band inversion, because of helical nature of the edge currents.

In HgTe quantum wells with band inversion, helical edge currents are mostly considered~\cite{zhang1,kane,zhang2} in the QSHE state, i.e. for bulk charge-neutrality  regime in zero magnetic field. However, the edge current has been demonstrated even to coexist with the conductive bulk in HgTe structures by a direct visualization technique~\cite{imaging}. It requires a low coupling between the edges and the bulk, i.e. due to the electrostatic depletion~\cite{shklovskii,image02} at the sample edge~\cite{imaging}.  From the continuous evolution of the edge current when the system is driven away from the charge-neutral regime, demonstrated in Ref.~\onlinecite{imaging}, we can assume that the edge current is still carried by the helical edge state, even for the conductive bulk. The coexistence is also possible from the theoretical considerations~\cite{pankratov}, in a crude similarity to the QH edge state transport in a dissipative regime between two neighbor QH plateaus~\cite{haug}. For our HgTe quantum wells, the helical edge states has been confirmed away from the charge-neutrality  regime by a direct calculation~\cite{raichev}.

%\subsection{Single F-2DEG junction}

In addition to the above considerations, we can experimentally demonstrate that the edge current is significant for transport to the  ferromagnetic side contacts in our samples. Examples of  $I-V$ characteristics are presented in Fig.~\ref{IVnl} for transport across a single junction N-2DEG (a) or F-2DEG (b,c). The $I-V$ curves are obtained in the experimental configuration presented in Fig.~\ref{sample} (b).

We use a reference normal junction N1 to verify the experimental setup. As it is expected for a standard three-point technique, $I-V$ curves coincide well for both potential probes $V_f$ and $V_b$, see Fig.~\ref{IVnl} (a).  Thus, the measured three-point $I-V$ curves in Fig.~\ref{IVnl} (a) reflect properties of the grounded contact N1. It is characterized by a high resistance (about 10~M$\Omega$), which indicates a significant depletion region at the 2DEG edge~\cite{shklovskii,image02}. 

In contrast, if we ground the permalloy ferromagnetic contact F1, the measured $I-V$ curve is crucially dependent on the mutual positions of current and voltage probes, see Fig.~\ref{IVnl} (b). We obtain significant signal $V_f$, i.e. for the voltage probe placed between the current and ground ones, but $V_b$ is always zero. Identical behavior is demonstrated in Fig.~\ref{IVnl} (c) for another ferromagnetic contact F2. We obtain the same behavior for both current polarities and for different current probes in Fig.~\ref{sample}, so the observed asymmetry between $V_f$ and $V_b$ is not connected with any absolute  direction in the sample. The asymmetry is only determined by  the mutual positions of the current and voltage contacts with respect to the grounded ferromagnetic lead. 

The asymmetry of the edge potential  always originates from a significant edge current contribution. It is obvious, e.g., in the conventional QH regime in high magnetic fields~\cite{buttiker,haug}. In the present experiment, the edge current is mostly flowing along the shortest (about 200~$\mu$m length) edge to the ground lead, because of the helical (counter-propagating) edge channel nature, in contrast to the conventional chiral QH case~\cite{buttiker}. Since the full length of the opposite mesa edge exceeds 2~mm (Fig.~\ref{sample} is not in scale), there should be no edge current flowing near the potential probe $V_b$, so the asymmetry between the potentials $V_f$ and $V_b$  is of a simple geometrical origin. Thus, the asymmetry in Fig.~\ref{IVnl} (b-c) is a direct experimental argument that the edge current is significant for transport to the  ferromagnetic side contacts in our samples.

The perfect ($V_b=0$ for any current $I$) asymmetry observed in Fig.~\ref{IVnl} (b-c) indicates that the grounded ferromagnetic side contact is strongly coupled to the conductive helical edge channel, while the reference Au one in Fig.~\ref{IVnl} (a)  demonstrates a standard low-coupling behavior. It can result from the specifically spin-dependent processes, like predicted in Refs.~\onlinecite{tanaka09,guigou}. We will discuss this strong coupling of the permalloy side contact in detail elsewhere.

%\section{Conclusion}

In a conclusion, we experimentally  investigate spin-polarized electron transport between two ferromagnetic contacts, placed at the edge of a two-dimensional electron system with band inversion. The system is realized in a narrow (8~nm) HgTe quantum well, the ferromagnetic side contacts are formed from a pre-magnetized permalloy film. In zero magnetic field,  we find a significant edge current contribution to the transport between two ferromagnetic contacts. We experimentally demonstrate that this transport is sensitive to the mutual orientation of the magnetization directions of two 200~$\mu$m-spaced ferromagnetic leads. This is a direct experimental evidence on the spin-coherent edge transport over the macroscopic  distances. Thus, the spin is extremely robust at the edge of a two-dimensional electron system with band inversion, confirming the helical spin-resolved nature of edge currents.

%\acknowledgments

We wish to thank V.T.~Dolgopolov, V.A.~Volkov, I.~Gornyi, and T.M.~Klapwijk for fruitful discussions.  We gratefully acknowledge financial support by the RFBR (projects No.~13-02-00065 and 13-02-12127) and RAS.

\end{document}